\documentclass[10pt, conference, compsocconf]{IEEEtran}

\usepackage{graphicx}
\usepackage{caption}
\usepackage{subfigure}

\usepackage{listings}
\usepackage{enumitem}
\setlist[itemize]{align=parleft,left=0pt..1em}

\usepackage{acronym}
\acrodef{ddp}{distributed data processing}
\acrodef{ddps}{\acl{ddp} system}
\acrodefplural{ddps}{\acl{ddp} systems}

\acrodef{RDD}{resilient distributed dataset}
\acrodefplural{RDD}{\acl{RDD}s}

\usepackage{comment}

\usepackage{booktabs}

\usepackage{tikz}
\usepackage{xcolor}
\newcommand*\circled[1]{\tikz[baseline=(char.base)]{
\node[shape=circle,fill,inner sep=0.5pt] (char) {\textcolor{white}{#1}};}}

\newcommand{\mask}[2]{\textbf{\textcolor{black}{#1}}}

\begin{document}
\title{Zero-Cost, Arrow-Enabled Data Interface for Apache Spark}

\author{\IEEEauthorblockN{
Sebastiaan Alvarez Rodriguez\IEEEauthorrefmark{1},
Jayjeet Chackrabroty\IEEEauthorrefmark{3},\\
Aaron Chu\IEEEauthorrefmark{3}, 
Ivo Jimenez\IEEEauthorrefmark{3},
Jeff LeFevre\IEEEauthorrefmark{3},
Carlos Maltzahn\IEEEauthorrefmark{3} and
Alexandru Uta\IEEEauthorrefmark{1}}
\IEEEauthorblockA{\IEEEauthorrefmark{1}LIACS, Leiden University\\
Email: sebastiaanalva@gmail.com, a.uta@liacs.leidenuniv.nl }
\IEEEauthorblockA{\IEEEauthorrefmark{3}UC Santa Cruz\\
Email: \{jchakra1,xweichu,ivotron,jlefevre,carlosm\}@ucsc.edu}}

\maketitle

\begin{abstract}
Distributed data processing ecosystems are widespread and their components are highly specialized, such that efficient interoperability is urgent. Recently, Apache Arrow was chosen by the community to serve as a format mediator, providing efficient in-memory data representation. Arrow enables efficient data movement between data processing and storage engines, significantly improving interoperability and overall performance. %Apache Spark, the de-facto data processing engine, cannot access data sources through the Arrow Dataset API and is unable to communicate efficiently with the ever growing group of Arrow-enabled frameworks. 
In this work, we design a new zero-cost data interoperability layer between Apache Spark and Arrow-based data sources through the Arrow Dataset API. Our novel data interface helps separate the computation (Spark) and data (Arrow) layers. This enables practitioners to seamlessly use Spark to access data from all Arrow Dataset API-enabled data sources and frameworks. To benefit our community, we open-source our work and show that consuming data through Apache Arrow is zero-cost: our novel data interface is either on-par or more performant than native Spark.
\end{abstract}

%\begin{IEEEkeywords}
% component; formatting; style; styling;
%distributed computing, computation offloading, optimization
%\end{IEEEkeywords}

% \fi
\IEEEpeerreviewmaketitle

\section{Introduction}
Distributed data processing frameworks, like Apache Spark~\cite{Spark}, Hadoop~\cite{Hadoop}, and Snowflake~\cite{Snowflake} have become pervasive, being used in most domains of science and technology.
The distributed data processing ecosystems are extensive and touch many application domains such as  
stream and event processing~\cite{Flink,Storm,Kafka,IBMEventStreams}, distributed machine learning~\cite{Ray}, or graph processing~\cite{Graphx}.
With data volumes increasing constantly, these applications are in urgent need of efficient interoperation through a common data layer format. In the absence of a common data interface, we identify two major problems: (1) data processing systems need to convert data, which is a very expensive operation; (2) data processing systems 
% need to implement
require new adapters or readers for each new data type to support and for each new system to integrate with.%, or whenever an integration with a new system is needed. 

A common example where these two issues occur is the de-facto standard data processing engine, Apache Spark. In Spark, the common data representation passed between operators is row-based~\cite{SparkSQL}. Connecting Spark to other systems such as MongoDB~\cite{MongoDB}, Azure SQL~\cite{Azure}, Snowflake~\cite{Snowflake}, or data sources such as Parquet~\cite{Parquet} or ORC~\cite{ORC}, entails building connectors and converting data. Although Spark was initially designed as a computation engine, this data adapter ecosystem was necessary to enable new types of workloads. However, we believe that using 
%the Arrow Dataset API
a universal interoperability layer
instead enables better and more efficient data processing, and more data-format related optimizations. 
% separating computation from the data interfacing layer enables better and more efficient interoperability, and more data-format related optimizations. 

The Arrow data format is available for many languages and is already adopted by many projects, including pySpark~\cite{Spark}, Dask~\cite{Dask}, Matlab~\cite{Matlab}, pandas~\cite{Python}, Tensorflow~\cite{Tensorflow}. 
Moreover, it is already used to exchange data between computation devices, such as CPUs and GPUs~\cite{CuDF}.
However, the Apache Arrow Dataset API~\cite{ArrowDatasetDocs}, not to be confused with the main Arrow~\cite{Arrow} library, 
emerged as a platform-independent data consumption API, which enables data processing frameworks to exchange columnar data efficiently, and without unnecessary conversions. 
% Arrow
The Arrow Dataset API
% aims to be the interoperability standard much sought after by all data processing frameworks. Through its APIs, Arrow
supports reading many kinds of datasources, both file formats and (remote) cloud storage. %It has the ability to move data between applications using extremely efficient in-memory columnar data representations.
Exploring the benefits of the Arrow Dataset API on building storage connectors is currently an understudied topic. 

In this paper, we therefore leverage the power of 
% Apache Arrow
the Apache Arrow Dataset API and separate the computation offered by Spark from the data (ingestion) layers, which are more efficiently handled by Arrow. We design a novel connector between Spark and 
%Arrow
The Apache Arrow Dataset API, to which Spark can offload its I/O. Using the Arrow Dataset API, we enable Spark access to Arrow-enabled data formats and sources. The increasing adoption of Arrow will make many more data types and sources available in the future, without adding any additional integration effort for our connector and, by extension, for Spark.
% Technically speaking, it would require: git merge. git push. That's it.

In this work, we lay the foundation of integrating Spark with all Arrow-enabled datasources and show that the performance achieved by our connector is promising, exceeding in many situations the performance achieved by Spark-implemented data connectors. We experiment with several design points, such as batch sizes, compression, data types (e.g., Parquet or CSV), and the scaling behavior of our connector. Our analysis shows that our proposed solution scales well, both with increasing data sizes and Spark cluster sizes, and we provide advice for practitioners on how to tune Arrow batch sizes and which compression algorithms to choose. Finally, practitioners can integrate our connector in existing programs without modification, since it is implemented as a drop-in, zero-cost replacement for Spark reading mechanisms. The contribution of this work is the following:
\begin{enumerate}[leftmargin=*]
    \item The design and implementation of a novel, zero-cost data interoperability layer between Apache Spark and 
    the Apache Arrow Dataset API. Our connector separates the computation (Spark) from the data (ingestion) layer (Arrow) and enables Spark interoperability with all Arrow-enabled formats and data sources. We open-source~\cite{arrowSparkAnonRepo} our implementation for the benefit of our community.
    \item The performance evaluation of the data interoperability layer. We show that Arrow-enabled Spark performs on-par or better than native-Spark and provide advice on how to tune Arrow parameters.
\end{enumerate}
\section{Design and Implementation}

\begin{figure}[tb]
    \centering
        \includegraphics[width=0.91\linewidth]{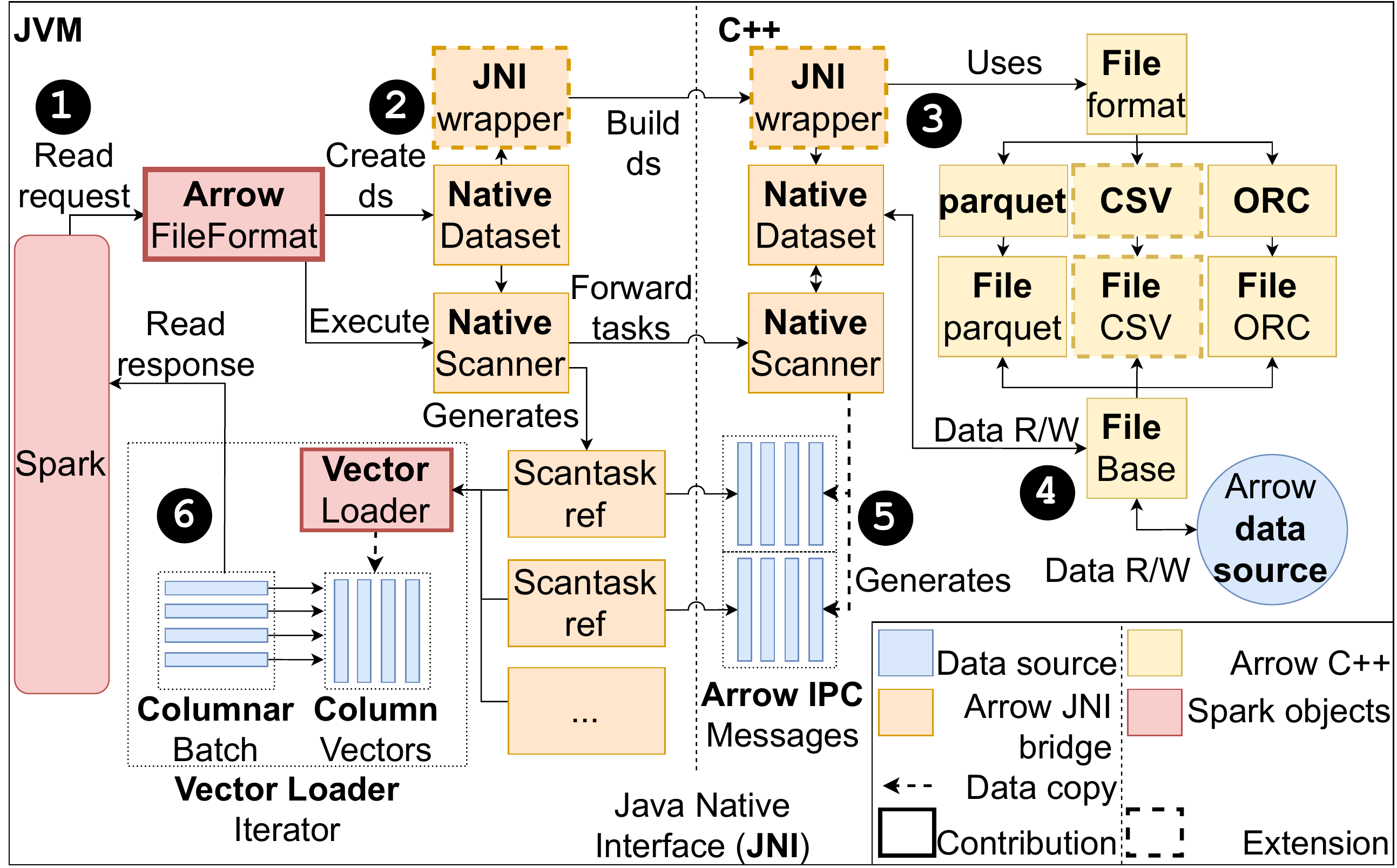}
          
    \caption{Arrow-Spark design overview, integrating Arrow (right) and Spark (left) using the Java Native Interface (JNI).} 
    \label{fig:overview_connector_arrow}
    \vspace*{-0.4cm}
\end{figure}

Our framework, called Arrow-Spark, provides an efficient interface between Apache Spark and all Arrow-enabled data sources and formats. Spark is in charge of execution, and Arrow provides the data, using its in-memory columnar formats. In Figure~\ref{fig:overview_connector_arrow}, we give a more detailed overview of how we access data through Arrow.
Spark is a JVM-based distributed data processing system, whereas the Arrow Dataset API~\cite{ArrowDatasetDocs} is only implemented in C++ and Python, but not Java. To enable communication, we created a bridge implementation.

\circled{\textbf{1}} \textbf{Reads.} Data transmission is initiated by a read request coming from Spark.
Read requests arrive at the \textit{Arrow FileFormat} datasource.

\circled{\textbf{2}} \textbf{JVM Dataset.} The \textit{Arrow FileFormat} constructs a \textit{Dataset} interface to read data through JNI, using the Arrow Dataset C++ API.
The JVM \textit{Dataset} interface forwards all calls through the \textit{JNI wrapper} to C++.
The Arrow \textit{Dataset} API interface is constructed on the C++-side, and a reference UUID is passed to the JVM interface counterpart.
Through this JVM interface, the \textit{Arrow FileFormat} 
% passes required information about datasources, and
initiates data scanning (reading) using a \textit{scanner}.

\circled{\textbf{3}} \textbf{Arrow C++ Dataset API.}
On the C++ side, a native \textit{Dataset} instance is created.
On creation, it picks a \textit{FileFormat}, depending on the type of data to be read.

\circled{\textbf{4}} \textbf{Data transmission.} The C++ Arrow Dataset API reads/writes the data in batches, using given FileFormat.

\circled{\textbf{5}} \textbf{Arrow IPC.} Each data batch is placed in memory as an Arrow IPC message, which is a columnar data format. The address to access each message is forwarded to the JVM, and stored in a \textit{Scantask} reference. Notice that here we make only one additional data copy. 

\circled{\textbf{6}} \textbf{Conversion.} Each Arrow IPC message is converted to an array of Spark-readable \textit{column vectors}.
Because Spark operators exchange row-wise data, we convert the \textit{column vectors} to a row-wise representation by wrapping the vectors in a \textit{ColumnarBatch}, which wraps columns and allows row-wise data access on them. This batch is returned to Spark and incurs data copying which cannot be avoided due to Spark operators only working on row-based data.
% There, row $x$ can be read by reading index $x$ of each vector.
%Finally, we return the \textit{ColumnarBatch}.

%\begin{comment}
When reading data in Arrow, one uses a Dataset object. Datasets contain required information about datasources, such as the location of the data. On the JVM side (e.g., in Spark), a Dataset object holds a references to a C++ Arrow Dataset. To obtain data, we scan the dataset using a Scanner. We execute the scanner, which provides us with a list of scan tasks. Each task loads the next batch of results into memory, in a columnar representation, as an Arrow Inter-Process Communication (IPC) message. To acquire the data as arrays of Spark-readable column vectors, we use a \textit{VectorSchemaRoot}. Once all necessary Arrow Vectors (columns) are extracted, we wrap the vectors in a \textit{ColumnarBatch}. As we tried to display in Figure~\ref{fig:overview_connector_arrow}, a ColumnarBatch provides an interface to read a group of columns in a row-by-row fashion. This is required, because Spark can only process row-wise data. Now Spark is able to iterate over rows of data, and this concludes the data provisioning through Arrow.

All interaction with Arrow is performed lazily, meaning they will only be executed on demand from Spark. This ensures we do not have to load in all data at once.
This is good for performance, because external datasources may be larger than the memory available on Spark clusters.
%\end{comment}

%\begin{comment}
\subsection{Usage, available APIs and extensions}
When designing our connector, we specifically searched for options to easily control Arrow from Scala, whether users need RDDs or the more modern Dataframes or Spark Datasets.
We show a simple example for loading RDDs, and another example for Dataframes.
\begin{figure}[tb]
    \centering
    \begin{lstlisting}[language=Scala,basicstyle=\scriptsize]
    def getAges: Unit {
        val conf: ArrowRDDReadConfig = 
                ArrowRDDReadConfig.builder()
                    .withPartitioner(...)
                    .withNumPartitions(100)
                    .withDataSourceURI(path)
                    .build()
        val context = session.sparkContext
        val peopleRDD: ArrowRDD = ArrowSpark.load(context, conf)
        peopleRDD.filter(row => row.getShort(1) > 42)
        // ...
    }
    \end{lstlisting}
    \caption{Example usage of our connector for Spark RDDs.}
    \label{fig:overview_usage_code_sample_rdd}
   % \todo[inline]{The code sample takes up much space, and does not fit. Remove?}
\end{figure}
In Figure~\ref{fig:overview_usage_code_sample_rdd}, we show how we can obtain a RDD.
We implemented a custom configuration object due to the many options that can be adjusted.
Users can configure simple options, like how many partitions are to be generated, but also extend the connector by providing a custom partitioner to be used when splitting the data, or even a custom object to load Arrow datasets. 
After an ArrowRDD is loaded, we can use it like any other RDD. Providing the file format is optional. If not specified, the connector will automatically determine what kind of data is requested. The example shows a filter operation on an RDD of individuals older than 42 years. 

\begin{figure}[tb]
    \centering
    \begin{lstlisting}[language=Scala,basicstyle=\scriptsize]
    def getAges: Unit {
        val df: DataFrame = session.read.arrow(path)
        df.createOrReplaceTempView("People")
        session.sql("SELECT name FROM People WHERE age > 42")
        // ...
    }
    \end{lstlisting}
    \caption{Example usage of our connector for Spark DataFrames, Spark Datasets}
    \label{fig:overview_usage_code_sample_modern}
    %\todo[inline]{The code sample takes up much space, and does not fit. Remove?}
\end{figure}

To show how interacting with our modern connector works, we provide an example in Figure~\ref{fig:overview_usage_code_sample_modern}. This example should be familiar to Spark users, as the common API is not changed. In ``session.read'', we obtain a DataFrameReader. The part right after, ``.arrow(path)'' calls an implicitly defined function, which loads the ``ArrowFileFormat'' object. This is all that is needed from a user-perspective to use this connector. The supplied path can point to either parquet- or CSV files. The ArrowFileFormat will automatically determine what kind of data is requested. %Like in the previous example, this example shows a query which filters people older than 42 years.
%\end{comment}

\subsection{Arrow-Spark JNI Bridge} 

Apache Spark (core) is implemented in Scala, and there does not exist an Arrow Dataset implementation written in any JVM language. 
The Arrow Dataset API~\cite{ArrowDatasetDocs} is not to be confused with main Arrow library, for which a Java stub implementation exists~\cite{ArrowJavaStub}.

\begin{comment}
Practitioners using Spark with Arrow are currently bound to a very small set of features, limited to PySpark, the Python implementation of Spark. These came out of the necessity to transform Python data to JVM data. Internally, PySpark is a wrapper for Python around core Spark. The Python bridge between Spark (JVM) and Arrow (C++) adds a highly inefficient link in applications, however, and requires developers to use only PySpark. The functionality limitations are related to only interfacing between Spark and \emph{pandas} data, as well as running UDFs on pandas dataframes. Arrow is then used to efficiently exchange data between these frameworks.
\end{comment}
Practitioners using Spark with Arrow are currently bound to a very small set of features. To use the pyarrow-dataset (Python wrappers around the C++ Arrow dataset API) implementation with PySpark (Python wrapper over Spark), one needs to implement these explicitly through the PySpark program, unlike our approach which is transparent to the programmer.
% These came out of the necessity to transform Python data to JVM data.
% Internally, PySpark is a wrapper for Python around core Spark.
Then, the Python bridge between Spark (JVM) and Arrow (C++) adds a highly inefficient link in applications, and a large functionality limitation.
% PySpark is not ready to be used with the Python Arrow dataset API wrapper, as it requires to convert the pyarrow dataset tables to a pyspark-readable format.
PySpark requires to convert the pyarrow dataset tables to \emph{pandas} data, a PySpark-readable format.
This conversion cancels Spark's lazy reading, and requires materializing the entire dataset into memory.
We experimented with a PySpark$+$pyarrow setup, and found it was consistently $30$ to $50$ times slower than our connector, with a growing performance difference when increasing dataset sizes.
Additionally, due to eager materialization dataset size is limited by RAM capacity.
% An additional disadvantage is that practitioners would need to write everything in Python for this to work.

Our work in this paper has a much wider scope, aiming to make core Spark read data using core Arrow, 
% removing inefficiencies/inefficient links,
providing universal data access to core Arrow from Core Spark and all its wrapper implementations at once.%in one go, using one library

Implementing our connector in core Spark (JVM) circumvents all aforementioned inefficiencies and shortcomings. We can therefore use our connector with all programming languages that Spark supports.

The core Arrow Dataset implementation is in native C++.
To access it, we use a Java Native Interface (JNI) implementation~\cite{JNI_arrow}, based on the Intel Optimized Analytics Platform project~\cite{intelbigdata}. Even though we could have chosen other languages for which there exists an Arrow Dataset implementation, we decided to use C++, because of its native-execution performance. Moreover, the Arrow Dataset API core is programmed in C++. Using any other language with Arrow bindings would add additional overhead. Even though the bridge between the JVM and native code brings a slight time penalty, we were able to minimize it by limiting data copies (see Figure~\ref{fig:overview_connector_arrow} for the  data copies that our interface implements). Additionally, C and C++ are the best-supported languages for interfacing with the JVM, through JNI. %In Figure~\ref{fig:overview_connector_arrow}, we see a detailed overview of how our connector works on the JVM- and C++ side.

\section{Arrow-Spark Performance}
We evaluate the performance of Arrow-Spark by comparing its performance with a standard Spark reading parquet and CSV data formats. %We adhere to reproducibility standards and describe the experimental environment and hardware. 
We performed $5$ separate experiments (E1-E5) to reveal various performance aspects of Arrow-Spark, as described in Table~\ref{tab:experiments}. Each experiment draws appropriate conclusions and provides advice for practitioners.

\begin{table}[]
\scriptsize
\centering
\caption{Description of the experiments in this work.}
\begin{tabular}{@{}lllll@{}}
\toprule
\textbf{Experiment} & \begin{tabular}[c]{@{}l@{}}Dataset\\ Size (GB)\end{tabular} & \begin{tabular}[c]{@{}l@{}}Batch\\ Size (KB)\end{tabular} & Query                                                                  & \begin{tabular}[c]{@{}l@{}}Row Size\\ (Bytes)\end{tabular} \\ \midrule
E1         & 1,144                                                         & 32-32,768                                                 & Scan: select *                                                         & $4 \times 8$                                                          \\
E2         & 71-4,500                                                     & 256                                                       & Scan: select *                                                         & $4 \times 8$                                                         \\
E3         & 71-1,144                                                     & 256                                                       & Scan: select *                                                         & $4 \times 8$                                                         \\
E4         & 71-1,144                                                     & 256                                                       & Scan: select *                                                         & $4 \times 8$                                                         \\
E5         & 715                                                         & 25,600                                                       & \begin{tabular}[c]{@{}l@{}}Projection:\\ select c1,c2,...\end{tabular} & $100 \times 8$                         \\ \bottomrule
\label{tab:experiments}
\vspace*{-0.7cm}
\end{tabular}
\end{table}

\begin{comment}
\begin{enumerate}[leftmargin=*]
    \item We measured how the data size influences reading time to determine how both systems scale when processing more data.
    \item We measured the influence of the cluster size on reading time to determine how both systems scale out.
    \item We experimented with different buffer sizes for our Arrow connector, to find out whether there are significant differences in reading times when Arrow has more or less buffer space to store intermediate data.
    \item We quantified how the two systems behave when reading different data sources: columnar (i.e., binary parquet format), and row-wise (text-based CSV format).
    \item We measured the influence of different compression techniques on various dataset sizes.
    %\item We verified how our Arrow-Spark system interacts with the Spark frontend by leveraging computation on RDDs, DataFrames, Spark SQL, and Spark Datasets.
    % \item We performed an experiment using a computational query. \sebastiaan{TODO: fill in once decided what query}
\end{enumerate}
%The general purpose for these 3 experiments is to get a comprehensive insight into the performance of our framework, and to show how this performance compares to using Spark's default reader.
\end{comment}

\subsection{Experiment Reproducibility}
% How I ran experiments and what reproducibility standards I followed
%We explain the general structure of all experiments. 
We used the specifications and parameters we give here as default values for every experiment, unless stated otherwise.%, we used parameters as described here.

\noindent\textbf{Experiment.} We ran every experiment 31 times for every configuration. We discard the first execution for every experiment, because the JVM- and memory caches are still `cold' during this run, producing an outlier. Between runs of the same experiment we did not close the Spark session as to keep JVM and caches warm. We adhered to the guidelines provided by Uta et al.~\cite{uta2020big}, to ensure our performance results are reproducible and significant. Due to the stability of our cluster, the difference between the $1st$ and $99th$ percentiles for each experiment were below 10\% (with one exception which we comment upon in Section~\ref{subsec:batchsizes}).

\noindent\textbf{Hardware and Deployment.} We ran all experiments on a cluster of 9 machines (8 Spark executor nodes and one driver), each equipped with 64\,GB RAM, and a dual 8-core Intel E5-2630v3 CPU running at 2.4\,GHz. Servers are interconnected with FDR InfiniBand connection ($\approx 56$\,Gb/s). %Each node has two 10,000\,rpm disks mounted in RAID-0. However, in our experiments, we do not use these drives as they are too slow and do not reveal the true I/O capabilities performance of either Spark or Arrow-Spark. We explain below how we experiment with only in-memory data. 
%We deployed Spark in cluster mode, and provide a Spark master node with its own physical server to operate on. We placed all Spark worker directories on local drives backed by disk. Additionally, we placed the input data to our application on RAMDisk (a memory-backed filesystem), to avoid the memory wall problem while experimenting with different file sizes.
%By default, we experiment on cluster size of 8 executor nodes, plus an additional node for the driver.
In the experiments, we always provide the number of nodes in terms of executor nodes. Spark is allowed to use up to $43$\,GB of the available $64$\,GB RAM in each node. Note that approximately $17$\,GB of RAM is already in use by input data as explained below. When reading parquet, we normally read uncompressed files unless stated otherwise.

\noindent\textbf{Dataset.} By default, we process $38.4\times10^9$ rows ($\approx1,144$\,GB) of synthetic data.
The data we experiment with has rows of $4$ up to $100$ columns which are 64-bit integers, split into files of $17$\,GB. During initial testing, we found that our experiments were influenced by local disk speed. With local disk reading, the bottleneck lies in disk speed, and we cannot uncover any framework inefficiencies. To mitigate this issue, we decided to deploy data as close as possible to the CPUs. We chose to deploy on RAMDisk, a memory-backed filesystem. Our machines have only 64\,GB RAM, while we want to experiment with datasets of much larger sizes. To solve this new problem, we created $X$ hardlinks for every file in a dataset of, e.g. $17$\,GB. Spark reads the hardlinks as if they were regular files, simulating a dataset size of $17\cdot(X+1)$\,GB. This way, we were able to experiment with virtually infinitely large datasets, regardless of the RAM size constraints.
%The integers of row $x$ are generated as $x>>3, x>>2, x>>1, x$, where `$>>$' denotes a Scala right shift operator. For our experiments, we generate $600,000,000$ rows of data.
%Depending on the chosen file format, this number of lines equals roughly $20$GB of data. In the experiments, we use much larger amounts of rows than our $64$GB data per node can hold. As explained in Section~\ref{sec:implementation_ram}, we want to store the whole dataset on RAMdisk, but datasets of e.g. $1280$GB cannot fit on $64$GB of RAM. We use $f-1$ hardlinks per file, to inflate $20$GB of data by a factor of $f$ without physically increasing memory usage. Note that we open-sourced our data generator, as well as all benchmarking code (see Section~\ref{sec:availability}). In several experiments, we read CSV files, or compressed parquet. We still generate $600,000,000$ rows for those configurations, and use hardlinks to reach the required dataset size on RAM.

\noindent\textbf{Performance Measurement.} We are interested in I/O performance. To measure it we created queries (see Table~\ref{tab:experiments}) which only read in all data, and count the number of rows as a way of triggering execution. This way, we can accurately measure the read performance. By default, we read this data into a Spark Dataframe (DF) to test with, reading $256$\,KB per batch. We replicated all experiments using Spark SQL, Datasets and directly RDDs, and the conclusions we drew were similar to what we present in this paper.

%\textbf{Final Words}
%For every experimental result, we precisely describe all relevant experiment parameters. This includes both unchanged default parameters and those parameters which we varied.  Finally, we made our benchmark and experimentation guidance software publicly available (see Section~\ref{sec:availability}). Every experiment parameter configuration can be found there, exactly the same as we used.

%\subsection{Experimentation on DRAM}\label{sec:implementation_ram}

\subsection{E1: Batch Size Tuning}\label{subsec:batchsizes}

\begin{figure}[tp]
    \centering
        \includegraphics[width=0.99\linewidth]{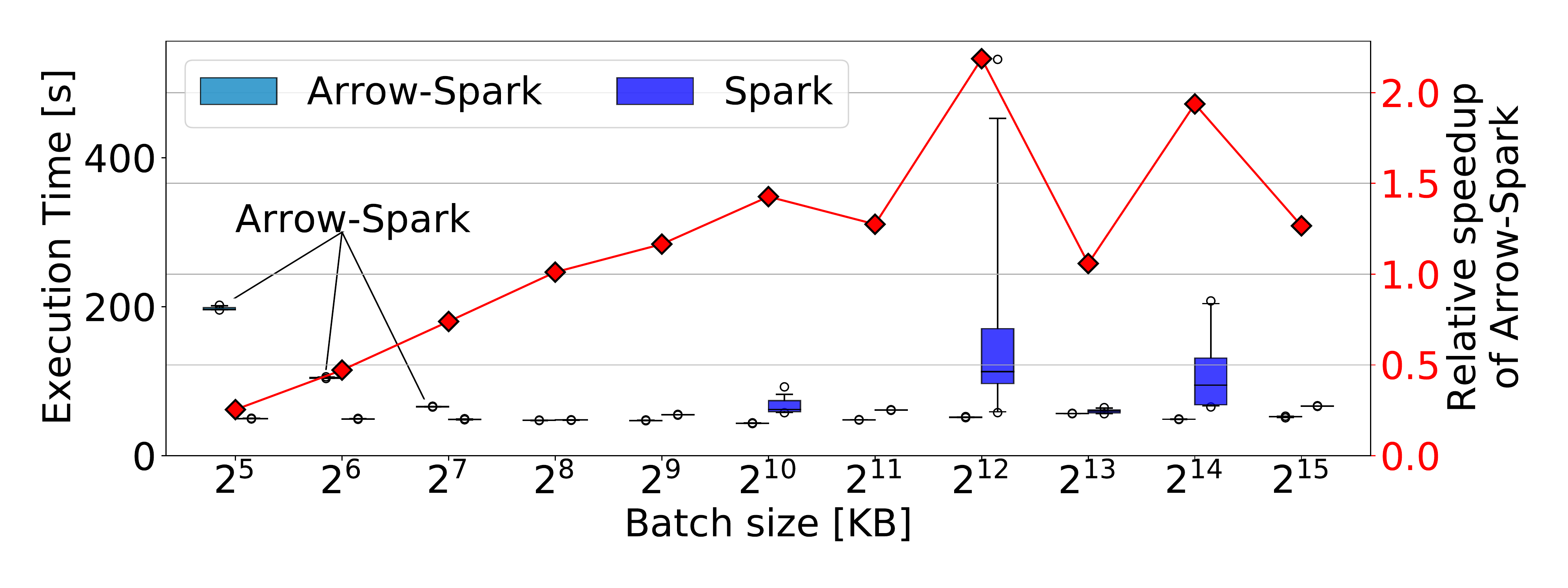}
          \vspace*{-0.5cm}
    \caption{\emph{Batch size matters:} Execution time (boxplots, left y-axis) and relative speedup (curve, \textcolor{red}{right y-axis}) with varying batch sizes.}
    \label{fig:res_buffersize_boxplot}
    \vspace*{-0.35cm}
\end{figure}

%We find that batch sizes are extremely important when performing I/O through Arrow. Practitioners should pay particular attention and do performance tuning on this parameter.
To request Arrow to read data, we place 
% columnar
(columnar) parquet data in memory buffers of limited sizes, i.e., \emph{batch sizes}. We measured the execution time under varying amounts of batch sizes. The results are plotted in Figure~\ref{fig:res_buffersize_boxplot}. The performance of default Spark remains largely unchanged when changing the amount of rows a buffer may hold for the smaller batch sizes. Only when setting this number to a very large amount, Spark seems to suffer a more significant overhead. After investigating this further, we found that the overhead is because Spark uses up all available memory with larger batch sizes, causing many garbage collection calls and memory swapping. Arrow is much more memory efficient, due to its use of streaming principles to transmit data.

For Arrow-Spark, choosing a batch size is much more important for performance. Low batch sizes degrade performance significantly (see first two boxplots  for $2^5$ and $2^6$ batch sizes). The root cause for this behaviour is the way Arrow-Spark works. For every batch, the Arrow Dataset API has to read data and transform it to IPC format. Further, the JVM-side reads this data and converts it to a Spark-understandable format. With smaller batches, there are more conversions and memory copies on both sides. %That is why Arrow-Spark becomes increasingly faster when using larger batch sizes, up to the  limit of $2^{18}$ rows. With such large batch sizes, claiming $8$\,MB of memory space per batch, we cannot efficiently use CPU caches anymore.%, and the performance we gain by using fewer but larger buffers does not compensate for memory cache misses from that point. 
Batches of 8192 rows offer the best performance with both systems, due to hardware features. Each row in our sample data consists of four $64$-bit integers. With 8192 rows, we get batches of exactly $256$\,KB. Our CPUs have a $256$\,KB L2 cache per core. (With different CPUs, the best batch size could be different.) By choosing a batch size of $8192$ rows, we can exactly fit one batch inside the L2-cache. We verified the correctness of this hypothesis by experimenting with other data shapes. %Due to size constraints, we skip these results in this paper.
%After batch sizes of $2^{13}$, default Spark becomes slightly slower. It does not have to do conversions, like Arrow-Spark, so it does not gain anything from using larger batch sizes. Using larger batch sizes than can fit in the L2 cache adds an additional layer of cache misses. This brings a slight reduction in performance.

\noindent \textbf{Conclusion-1:} Practitioners should always overestimate rather than underestimate batch sizes for Arrow-Spark. Underestimations cause strong performance degradation.% , while overestimations only add relatively little overhead. 

\subsection{E2: Data Scalability}
\begin{figure}[t]
     \centering
     \includegraphics[width=0.99\linewidth]{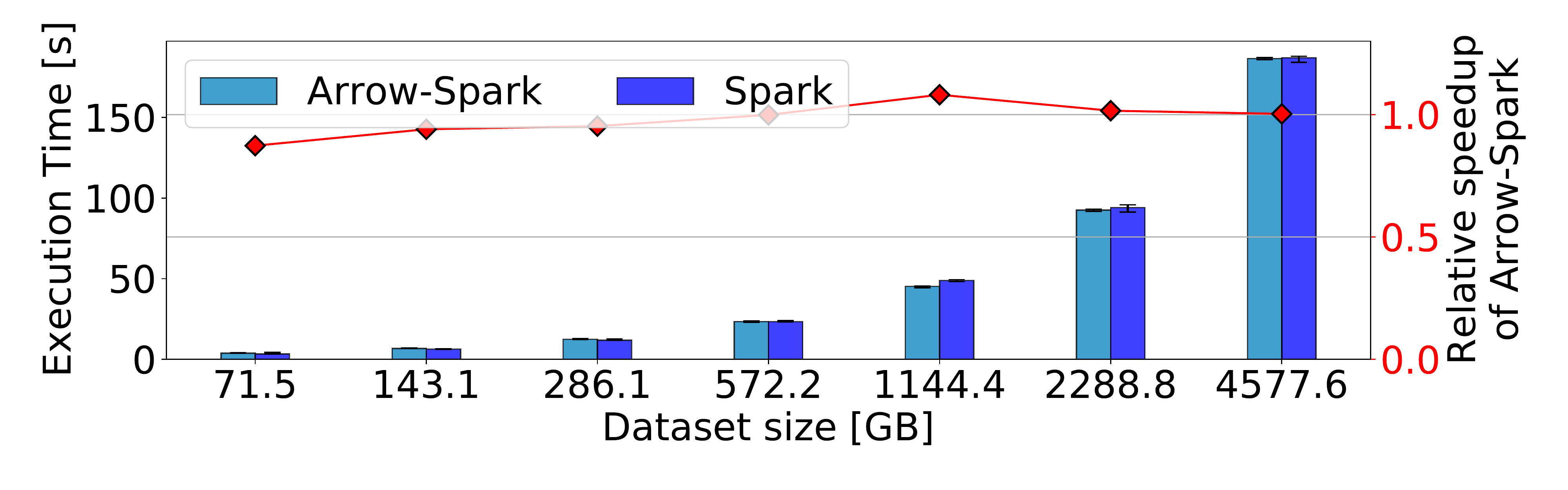}
       \vspace*{-0.5cm}
    \caption{\emph{Arrow-Spark performance is good:} Execution time (bars) and relative speedup (curve) on increasing parquet data sets.}
    \label{fig:res_data_scalability_boxplot}
    \vspace*{-0.55cm}
\end{figure}

%Arrow-Spark scales well with increasing dataset sizes. 
An important property of a \acl{ddps} is its \emph{data scalability} behavior, which evaluates system performance with increasing dataset sizes. %When performance scales (almost) linearly with dataset sizes, we can process extremely large quantities of data with minimal overheads.
We measured data scalability by reading parquet datasets with a wide range of different sizes. The results of this experiment are depicted in Figure~\ref{fig:res_data_scalability_boxplot}. This figure plots the execution time with increasing dataset sizes (left vertical axis), as well as the relative speedup (or slowdown) between the two systems (right vertical axis). The execution time approximately doubles as the dataset size doubles. Both Spark and Arrow-Spark scale well. %with dataset size.

Despite being slower than Spark on the smaller datasets we tested, Arrow-Spark gradually becomes relatively faster than Spark for larger datasets. We depict this effect in the lineplot of Figure~\ref{fig:res_data_scalability_boxplot}, which shows the relative speedup of our framework, when compared to default Spark. This effect is because Arrow-Spark has several sources of constant overhead. Using the JNI bridge is the largest cause of overhead. The reading itself is slightly faster than with Spark, causing the difference between the measured systems to grow with the increase in dataset size. We found similar patterns when scaling the cluster size between 4 and 32 nodes. %Due to space constraints, we were unable to place these experiments in this paper.

\noindent \textbf{Conclusion-2:} Arrow-Spark scales well with dataset sizes. Its advantage over Spark is highest with 1-2\,TB datasets. Practitioners can leverage Arrow-Spark at zero additional cost with larger dataset sizes.

\subsection{E3: Row-wise Formats}

\begin{figure}[tp]
    \centering
    \includegraphics[width=0.99\linewidth]{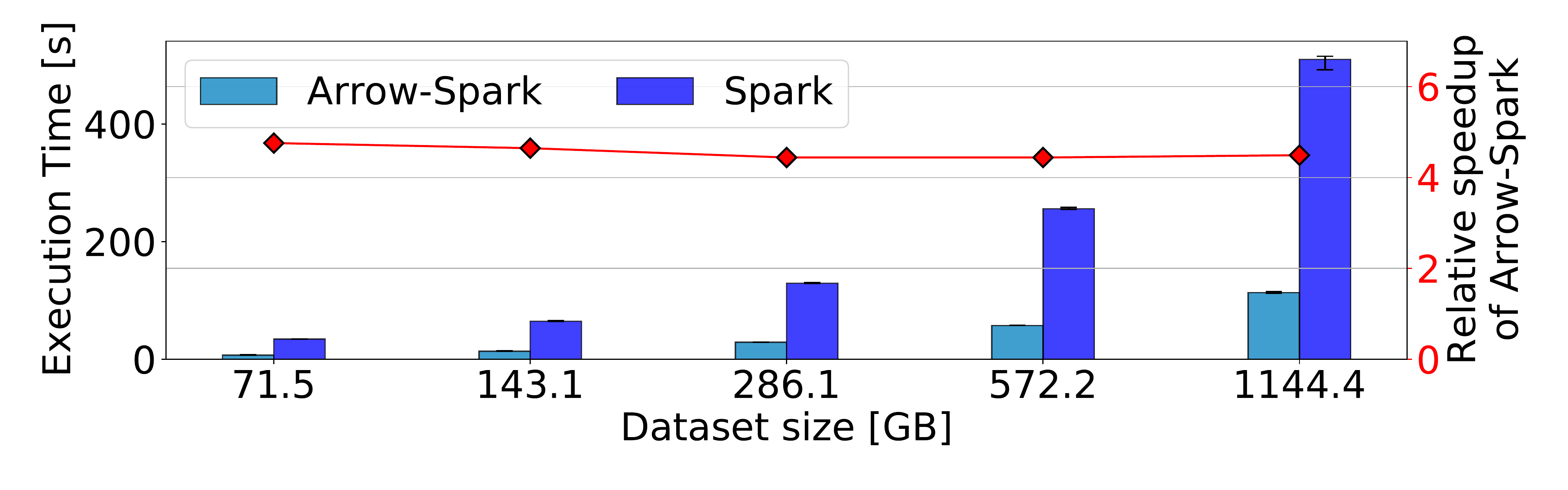}
      \vspace*{-0.5cm}
    \caption{\emph{Arrow-Spark is faster than Spark with CSV files:} Execution time (bars) and relative speedup (curve) on increasing CSV data sets.}
    \label{fig:res_row_vs_columnar_boxplot}
   \vspace*{-0.55cm}
\end{figure}

Arrow-Spark supports multiple file formats through Arrow functionality. We performed an experiment comparing Arrow-Spark and default Spark on CSV data. The results are displayed in Figure~\ref{fig:res_row_vs_columnar_boxplot}. Our implementation is significantly faster than Spark, starting at a speedup factor of approximately $4.5$. Larger datasets do not influence this speedup, which remains constant over all dataset sizes we tested.

This is due to Spark's inefficient CSV reader, the Java-based Univocity-CSV parser~\cite{univocity}. By contrast, our connector works with the highly efficient C++ Arrow Dataset API implementation. In cases where parsing is involved, using native code commonly is much more performant than other solutions. The cost of processing more data is higher for Spark, and much lower for Arrow-Spark due to the differences in parsers. This explains why using larger datasets results in a relatively bigger runtime increase for Spark. 

\noindent \textbf{Conclusion-3:} Overall, Arrow-Spark brings significantly higher performance for ingesting text-based file formats than default Spark and the performance difference is constant with respect to dataset size. Practitioners should always choose Arrow-Spark for such file formats.

\subsection{E4: Parquet Compression}
% \begin{figure}[tb]
%     \makebox[\textwidth][c] {
%         \includegraphics[width=\linewidth]{fig/results/compression/boxplot_compression_arrow.pdf}
%     }
%     \caption{Measured execution time for Arrow-Spark on varying amounts of input rows, for a cluster of $8$ executor nodes, reading from uncompressed, gzip-compressed and snappy-compressed parquet files in batches of $8192$ rows, using Dataframes.}
%     \label{fig:res_compression_arrow_boxplot}
%     \vspace{-1cm}
% \end{figure}
% \begin{figure}[tb]
%     \makebox[\textwidth][c] {
%         \includegraphics[width=\linewidth]{fig/results/compression/boxplot_compression_default.pdf}
%     }
%     \caption{Measured execution time for default Spark on varying amounts of input rows, for a cluster of $8$ executor nodes, reading from uncompressed, gzip-compressed and snappy-compressed parquet files in batches of $8192$ rows, using Dataframes.}
%     \label{fig:res_compression_spark_boxplot}
%     \vspace{-1.5cm}
% \end{figure}

\begin{comment} % Old figure for both Spark and Arrow-Spark
\begin{figure}[tp]
\centering
\subfigure[Measured execution time for Arrow-Spark.]{
    \label{fig:res_compression_arrow_boxplot}
    \includegraphics[width=0.9\linewidth]{fig/results/compression/boxplot_compression_arrow.pdf}
}
\qquad
\subfigure[Measured execution time for default Spark.]{
\label{fig:res_compression_spark_boxplot}
  \includegraphics[width=0.45\linewidth]{fig/results/compression/boxplot_compression_default.pdf}
}
\caption{Measured execution time for Arrow-Spark on varying amounts of input rows.}
\end{figure}
\end{comment}

\begin{figure}[tp]
    \centering
    \includegraphics[width=0.91\linewidth]{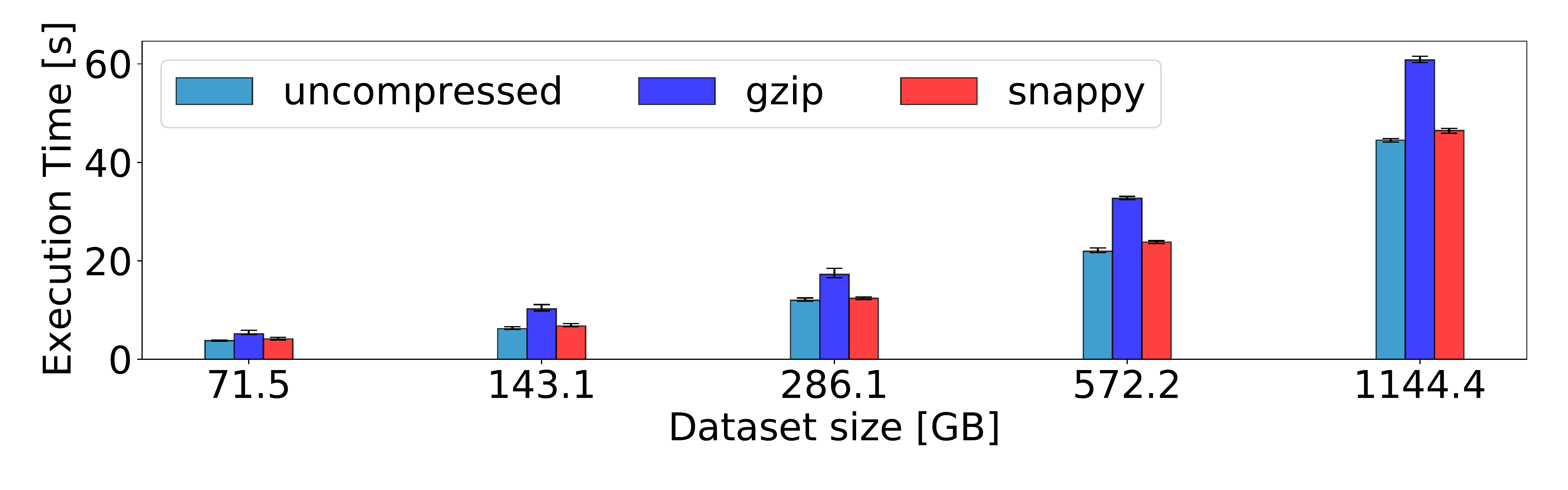}
    \vspace*{-0.2cm}
    \caption{\emph{Arrow-Spark leverages compression:} Execution time for Arrow-Spark with different compression techniques for parquet data.}
    \label{fig:res_compression_arrow_boxplot}
    \vspace*{-0.3cm}
\end{figure}

Arrow-Spark can leverage parquet compression. We compare the performance of Arrow-Spark under several compression options. The results can be found in Figure~\ref{fig:res_compression_arrow_boxplot}. Note that using compressed parquet files in practical situations produces vastly different results from the results we obtained in our experimental setup using RAMDisk. Usually, the bottleneck for reading data is I/O. Using compressed data trades I/O for increased CPU load. In our setup, we deploy data on RAMDisk, and we have no I/O bottleneck. Reading compressed data only increases CPU load in this case, decreasing performance. We compare reading \emph{uncompressed} parquet with \emph{snappy} and \emph{gzip}. When reading from memory, without much I/O overhead, reading compressed data is slower due to the added CPU overhead of decompressing it.

\noindent \textbf{Conclusion-4:} Arrow-Spark is able to leverage various compression algorithms for parquet files. When leveraging slower I/O media, these can help practitioners by trading I/O volume for increased CPU-load.

\subsection{E5: Column Projection}

\begin{figure}[tp]
    \centering
    \includegraphics[width=0.99\linewidth]{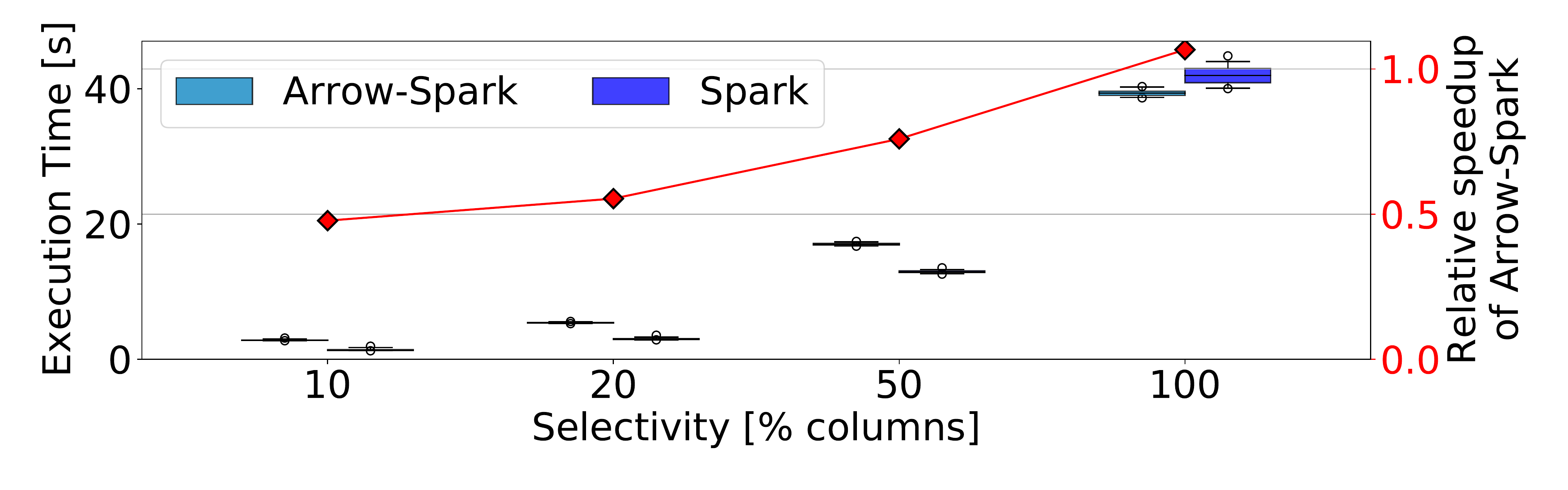}
    \vspace*{-0.5cm}
    \caption{\emph{Arrow-Spark pushes down projections:} Execution time (boxplots) and relative speedup (curve) when reading 9.6 billion rows in batches of 32,768 rows, varied projection selectivity.}
    \label{fig:res_projection_boxplot}
   \vspace*{-0.45cm}
\end{figure}

%Arrow-Spark is able to efficiently leverage columnar formats for projections.
One benefit of many columnar dataformats is the ability to project columns, i.e. to select a subset of the total number of columns for reading at little to no cost.
Arrow-Spark pushes down projection operations from Spark to Arrow.
The benefit of doing this is that 
Arrow becomes in charge of providing subsets of data. %In itself, Arrow is much more memory-efficient than reading through Spark-provided methods, because Arrow uses in-memory datasets, providing data through streaming patterns. 
Practitioners need not change code, but use the Arrow-Spark interface we designed, because projections are directly controlled by the Spark query optimizer. The results can be found in Figure~\ref{fig:res_projection_boxplot}.
When selecting very few columns, Spark is relatively quicker because it 
% also pushes down projections to its \emph{parquet} reader which basically skips data reading. 
pushes down projections to its parquet reader, while Arrow-Spark pushes down to Arrow over the JNI.
% pushes down projections directly to parquet, while Arrow-Spark pushes down to Arrow over the JNI bridge.
When selecting more columns, Arrow-Spark becomes relatively faster due to its advantage over larger data sizes. Overall, this behavior is consistent with findings from E2. Although Spark is seemingly faster at higher selectivities, this is relative: on real-world datasets, even 1\% selectivity could leverage datasets of over 1TB, at which point Arrow-Spark's performance is superior.

\noindent \textbf{Conclusion-5:} Arrow-Spark is able to leverage projections for parquet files, effectively pushing down projection queries to the data layer, minimizing data movement.

\section{Related Work}
%We review related work in two categories: %(1) Connector extensions for Spark, and (2) Decoupling execution from data ingestion.

%\noindent\textbf{Connector Extensions for Spark.} 
Several connector extensions allow users to read new file formats or datasource backends, such as HDF5 or netCDF~\cite{liu2016h5spark}, as well as extensions on Hadoop to allow efficiently reading of highly structured array-based data. Albis~\cite{Albis}, proposes a new columnar-hybrid format to be used with modern hardware. Frameworks like MongoSpark~\cite{MongoDB_Spark} and Databricks Spark-RedShift~\cite{databricks_redshift} connect a specific backend to Spark. There also are extensions to improve existing reading implementations: Intel Optimized Analytics Package (OAP)~\cite{intelbigdata}, and the JVM Arrow dataset API~\cite{JNI_arrow}. Our connector has a different approach to memory allocation, backward-compatibility, and we support more file types at the moment.
%\textcolor{red}{\textbf{TODO}: Need to explain here that (1.) We just used their stuff, combined it \& improved it, (2.) We did a whole bunch of experiments to evaluate performance and write a paper, while they evaluated\&published nothing.}

% \textbf{Universal Data Layers.}
% The Arrow Dataset API~\cite{Arrow} is an in-memory database, which aims to be the de-facto interoperability layer between data processing frameworks and datasources.
% Li et al.~\cite{li2020mainlining} make the case for using in-memory datasets in Database Management Systems to greatly reduce the time needed to export data to Distributed Data Processing Systems.
%\noindent\textbf{Decoupling Execution from Data Ingestion.}
%In this work, we made the case for separate execution and data ingestion.
%We were not the first to split data and execution.
%When the data is only available from remote datasources, we already have some sort of a separation. 
%Databricks~\cite{databricks}, a provider cloud resources involving Spark, Delta Lake, etc, provides several interesting connectors, 
Databricks~\cite{databricks} also provides several connectors separating execution from data layers, such as JDBC and ODBC connectors~\cite{databricks_jdbc}. %These connectors allows users to query many MySQL-capable backends.
The difference between such projects and our work is that we provide a separation layer between execution and data ingestion with high interoperability for any (Arrow-enabled) datasource. There is no official support for using Core (Scala) Spark together with Core Arrow (C++). Our system provides such support.
% Decoupling execution and data ingestion on one cluster is something else, however.
% While the goal of the aforementioned connectors is to allow users to obtain data from remote services, our goal is to improve performance on the way.
% Actually, we were unable to find scientific work about decoupling execution and data and experimenting with it.
%We were unable to find scientific work about decoupling execution and data ingestion. In Spark+AI Summit 2020, V. Ganesh~\cite{sparksummit2020_arrow} speaks very briefly about using Arrow for reading data through some Python libraries. There is no official support for using Core (Scala) Spark together with Core Arrow (C++). Our system provides such support.
\section{Conclusion}
%In this work, we aimed to answer the following questions:
%\begin{itemize}
%    \item[RQ1] How to build a communication bridge between Apache Arrow and Apache Spark, allowing Spark to access data through Arrow?
%    \item[RQ2] How does Spark performance compare when using our Arrow support implementation versus using Spark's built-in support? 
    % / What is the relative performance between...
%\end{itemize}

%To answer these questions, we built a connector between core Spark and core Arrow, to separate execution from data ingestion.
%We carefully explain how we approached construction of this connector, and made it open-source, to benefit the community.
We investigated decoupling the computation and the data (ingestion) layers. This is needed for enabling interoperability and reducing the amount of data conversion. We built a prototype that leverages Arrow-enabled data sources to Apache Spark, effectively decoupling Spark's computation from Arrow's data ingestion. Through our experimentation, we concluded that using connectors like ours is zero-cost: Arrow-Spark not only
% does not slow down overall performance, but in some cases even significantly improves it.
% improves performance in some cases, but also shows similarly good performance to Spark in most cases.
retains overall performance, but in some cases even significantly improves it.
Next to improved performance, we gain the ability to access any datasource that implements support for Arrow, allowing us to connect to many different data processing frameworks, data storage systems and file formats.

\section*{Acknowledgements}
The work in this article was in part supported by The Dutch National Science Foundation NWO Veni grant VI.202.195, by the US National Science Foundation under Cooperative Agreement OAC-1836650, by the US Department of Energy ASCR DE-NA0003525 (FWP 20-023266), and by the Center for Research in Open Source Software (cross.ucsc.edu)

\bibliographystyle{style/IEEEtran}
\bibliography{bib/main,bib/related,bib/frameworks}

\begin{thebibliography}{10}
\providecommand{\url}[1]{#1}
\csname url@samestyle\endcsname
\providecommand{\newblock}{\relax}
\providecommand{\bibinfo}[2]{#2}
\providecommand{\BIBentrySTDinterwordspacing}{\spaceskip=0pt\relax}
\providecommand{\BIBentryALTinterwordstretchfactor}{4}
\providecommand{\BIBentryALTinterwordspacing}{\spaceskip=\fontdimen2\font plus
\BIBentryALTinterwordstretchfactor\fontdimen3\font minus
  \fontdimen4\font\relax}
\providecommand{\BIBforeignlanguage}[2]{{%
\expandafter\ifx\csname l@#1\endcsname\relax
\typeout{** WARNING: IEEEtran.bst: No hyphenation pattern has been}%
\typeout{** loaded for the language `#1'. Using the pattern for}%
\typeout{** the default language instead.}%
\else
\language=\csname l@#1\endcsname
\fi
#2}}
\providecommand{\BIBdecl}{\relax}
\BIBdecl

\bibitem{Spark}
M.~Zaharia, M.~Chowdhury, M.~J. Franklin, S.~Shenker, I.~Stoica \emph{et~al.},
  ``Spark: Cluster computing with working sets.'' \emph{HotCloud}, vol.~10, no.
  10-10, p.~95, 2010.

\bibitem{Hadoop}
\BIBentryALTinterwordspacing
{Apache Software Foundation}, ``Hadoop.'' [Online]. Available:
  \url{https://hadoop.apache.org}
\BIBentrySTDinterwordspacing

\bibitem{Snowflake}
B.~Dageville, T.~Cruanes, M.~Zukowski, V.~Antonov, A.~Avanes, J.~Bock,
  J.~Claybaugh, D.~Engovatov, M.~Hentschel, J.~Huang \emph{et~al.}, ``The
  snowflake elastic data warehouse,'' in \emph{Proceedings of the 2016
  International Conference on Management of Data}, 2016, pp. 215--226.

\bibitem{Flink}
P.~Carbone, A.~Katsifodimos, S.~Ewen, V.~Markl, S.~Haridi, and K.~Tzoumas,
  ``Apache flink: Stream and batch processing in a single engine,''
  \emph{Bulletin of the IEEE Computer Society Technical Committee on Data
  Engineering}, vol.~36, no.~4, 2015.

\bibitem{Storm}
M.~H. Iqbal and T.~R. Soomro, ``Big data analysis: Apache storm perspective,''
  \emph{International journal of computer trends and technology}, vol.~19,
  no.~1, pp. 9--14, 2015.

\bibitem{Kafka}
R.~{Shree}, T.~{Choudhury}, S.~C. {Gupta}, and P.~{Kumar}, ``Kafka: The modern
  platform for data management and analysis in big data domain,'' in \emph{2017
  2nd International Conference on Telecommunication and Networks (TEL-NET)},
  2017, pp. 1--5.

\bibitem{IBMEventStreams}
M.~Hirzel, H.~Andrade, B.~Gedik, G.~Jacques-Silva, R.~Khandekar, V.~Kumar,
  M.~Mendell, H.~Nasgaard, S.~Schneider, R.~Soul{\'e} \emph{et~al.}, ``Ibm
  streams processing language: Analyzing big data in motion,'' \emph{IBM
  Journal of Research and Development}, vol.~57, no. 3/4, pp. 7--1, 2013.

\bibitem{Ray}
P.~Moritz, R.~Nishihara, S.~Wang, A.~Tumanov, R.~Liaw, E.~Liang, M.~Elibol,
  Z.~Yang, W.~Paul, M.~I. Jordan, and I.~Stoica, ``Ray: A distributed framework
  for emerging ai applications,'' in \emph{Proceedings of the 13th USENIX
  Conference on Operating Systems Design and Implementation}, ser.
  OSDI'18.\hskip 1em plus 0.5em minus 0.4em\relax USA: USENIX Association,
  2018, p. 561–577.

\bibitem{Graphx}
R.~S. Xin, J.~E. Gonzalez, M.~J. Franklin, and I.~Stoica, ``Graphx: A resilient
  distributed graph system on spark,'' in \emph{First international workshop on
  graph data management experiences and systems}, 2013, pp. 1--6.

\bibitem{SparkSQL}
M.~Armbrust, R.~S. Xin, C.~Lian, Y.~Huai, D.~Liu, J.~K. Bradley, X.~Meng,
  T.~Kaftan, M.~J. Franklin, A.~Ghodsi \emph{et~al.}, ``Spark sql: Relational
  data processing in spark,'' in \emph{Proceedings of the 2015 ACM SIGMOD
  international conference on management of data}, 2015, pp. 1383--1394.

\bibitem{MongoDB}
K.~Chodorow, \emph{MongoDB: the definitive guide: powerful and scalable data
  storage}.\hskip 1em plus 0.5em minus 0.4em\relax " O'Reilly Media, Inc.",
  2013.

\bibitem{Azure}
D.~Chappell \emph{et~al.}, ``Introducing windows azure,'' \emph{Microsoft, Inc,
  Tech. Rep}, 2009.

\bibitem{Parquet}
D.~Vohra, ``Apache parquet,'' in \emph{Practical Hadoop Ecosystem}.\hskip 1em
  plus 0.5em minus 0.4em\relax Springer, 2016, pp. 325--335.

\bibitem{ORC}
Apache, ``Apache orc - high-performance columnar storage for hadoop,''
  \url{https://orc.apache.org/}, Apache Software Foundation, accessed:
  2020-11-06.

\bibitem{Dask}
M.~Rocklin, ``Dask: Parallel computation with blocked algorithms and task
  scheduling,'' in \emph{Proceedings of the 14th python in science conference},
  vol. 126.\hskip 1em plus 0.5em minus 0.4em\relax Citeseer, 2015.

\bibitem{Matlab}
MATLAB, \emph{version 7.10.0 (R2010a)}.\hskip 1em plus 0.5em minus 0.4em\relax
  Natick, Massachusetts: The MathWorks Inc., 2010.

\bibitem{Python}
G.~Van~Rossum \emph{et~al.}, ``Python,'' 1991.

\bibitem{Tensorflow}
M.~Abadi, P.~Barham, J.~Chen, Z.~Chen, A.~Davis, J.~Dean, M.~Devin,
  S.~Ghemawat, G.~Irving, M.~Isard \emph{et~al.}, ``Tensorflow: A system for
  large-scale machine learning,'' in \emph{12th $\{$USENIX$\}$ symposium on
  operating systems design and implementation ($\{$OSDI$\}$ 16)}, 2016, pp.
  265--283.

\bibitem{CuDF}
cuDF community, ``cudf - gpu dataframes,''
  \url{https://github.com/rapidsai/cudf}, RAPIDS, accessed: 2020-11-16.

\bibitem{ArrowDatasetDocs}
A.~D. Team,
  \url{https://arrow.apache.org/docs/python/dataset.html\#reading-from-cloud-storage},
  Apache Software Foundation, accessed: 2020-11-08.

\bibitem{Arrow}
------, ``Apache arrow,'' \url{https://arrow.apache.org}, 10 2018.

\bibitem{arrowSparkAnonRepo}
Anonymous, ``Arrow-spark,'' \frameworkrepo, 2021.

\bibitem{ArrowJavaStub}
A.~D. Team, ``Apache arrow java implementation,''
  \url{https://arrow.apache.org/docs/java/}, 10 2018.

\bibitem{JNI_arrow}
H.~Zhang, ``Arrow-7808: [java][dataset] implement dataset java api by jni to
  c++,'' \url{https://github.com/zhztheplayer/arrow-1/tree/ARROW-7808}, Github,
  Apache Arrow community, accessed: 2020-09-14.

\bibitem{intelbigdata}
I.~Corporation, ``Optimized analytics package for spark platform (oap),''
  \url{https://github.com/Intel-bigdata/OAP }, Github, Intel Corporation,
  accessed: 2021-01-15.

\bibitem{uta2020big}
A.~Uta, A.~Custura, D.~Duplyakin, I.~Jimenez, J.~Rellermeyer, C.~Maltzahn,
  R.~Ricci, and A.~Iosup, ``Is big data performance reproducible in modern
  cloud networks?'' in \emph{17th $\{$USENIX$\}$ Symposium on Networked Systems
  Design and Implementation ($\{$NSDI$\}$ 20)}, 2020, pp. 513--527.

\bibitem{univocity}
T.~univocity team, ``univocity-parsers,''
  \url{https://github.com/uniVocity/univocity-parsers }, Github, uniVocity,
  accessed: 2021-02-01.

\bibitem{liu2016h5spark}
J.~Liu, E.~Racah, Q.~Koziol, R.~S. Canon, and A.~Gittens, ``H5spark: bridging
  the i/o gap between spark and scientific data formats on hpc systems,''
  \emph{Cray user group}, 2016.

\bibitem{Albis}
A.~Trivedi, P.~Stuedi, J.~Pfefferle, A.~Schuepbach, and B.~Metzler, ``Albis:
  High-performance file format for big data systems,'' in \emph{2018 USENIX
  Annual Technical Conference (USENIX ATC 18)}, 2018, pp. 615--630.

\bibitem{MongoDB_Spark}
R.~Lawley, ``Apache spark connector for mongodb,''
  \url{https://www.slideshare.net/mongodb/how-to-connect-spark-to-your-own-datasource},
  \url{https://databricks.com/blog/2015/03/20/using-mongodb-with-spark.html},
  MongoDB, accessed: 2020-11-06.

\bibitem{databricks_redshift}
Databricks, ``Databricks,'' \url{https://github.com/databricks/spark-redshift},
  Databricks, Github, accessed: 2021-02-01.

\bibitem{databricks}
------, ``Databricks,'' \url{https://databricks.com/}, Databricks, accessed:
  2021-02-01.

\bibitem{databricks_jdbc}
------, ``Databricks,''
  \url{https://docs.databricks.com/data/data-sources/sql-databases.html},
  Databricks, accessed: 2021-02-01.

\end{thebibliography}

\end{document}